\input amssym.def
\input epsf

% Page layout

\magnification=\magstephalf
\hsize=14.0 true cm
\vsize=19 true cm
\hoffset=1.0 true cm
\voffset=2.0 true cm

\abovedisplayskip=12pt plus 3pt minus 3pt
\belowdisplayskip=12pt plus 3pt minus 3pt
\parindent=1.0em

% Fonts

\font\sixrm=cmr6
\font\eightrm=cmr8
\font\ninerm=cmr9

\font\sixi=cmmi6
\font\eighti=cmmi8
\font\ninei=cmmi9

\font\sixsy=cmsy6
\font\eightsy=cmsy8
\font\ninesy=cmsy9

\font\sixbf=cmbx6
\font\eightbf=cmbx8
\font\ninebf=cmbx9

\font\eightit=cmti8
\font\nineit=cmti9

\font\eightsl=cmsl8
\font\ninesl=cmsl9

\font\sixss=cmss8 at 8 true pt
\font\sevenss=cmss9 at 9 true pt
\font\eightss=cmss8
\font\niness=cmss9
\font\tenss=cmss10

 at 12 true pt
\font\bigrm=cmr10 at 12 true pt
 at 12 true pt

 at 14 true pt
\font\Bigrm=cmr12 at 16 true pt
 at 14 true pt

\catcode`@=11
\newfam\ssfam

\def\tenpoint{\def\rm{\fam0\tenrm}%
    \textfont0=\tenrm \scriptfont0=\sevenrm \scriptscriptfont0=\fiverm
    \textfont1=\teni  \scriptfont1=\seveni  \scriptscriptfont1=\fivei
    \textfont2=\tensy \scriptfont2=\sevensy \scriptscriptfont2=\fivesy
    \textfont3=\tenex \scriptfont3=\tenex   \scriptscriptfont3=\tenex
    \textfont\itfam=\tenit                  \def\it{\fam\itfam\tenit}%
    \textfont\slfam=\tensl                  \def\sl{\fam\slfam\tensl}%
    \textfont\bffam=\tenbf \scriptfont\bffam=\sevenbf
    \scriptscriptfont\bffam=\fivebf
                                            \def\bf{\fam\bffam\tenbf}%
    \textfont\ssfam=\tenss \scriptfont\ssfam=\sevenss
    \scriptscriptfont\ssfam=\sevenss
                                            \def\ss{\fam\ssfam\tenss}%
    \normalbaselineskip=13pt
    \setbox\strutbox=\hbox{\vrule height8.5pt depth3.5pt width0pt}%
    \let\big=\tenbig
    \normalbaselines\rm}

\def\ninepoint{\def\rm{\fam0\ninerm}%
    \textfont0=\ninerm      \scriptfont0=\sixrm
                            \scriptscriptfont0=\fiverm
    \textfont1=\ninei       \scriptfont1=\sixi
                            \scriptscriptfont1=\fivei
    \textfont2=\ninesy      \scriptfont2=\sixsy
                            \scriptscriptfont2=\fivesy
    \textfont3=\tenex       \scriptfont3=\tenex
                            \scriptscriptfont3=\tenex
    \textfont\itfam=\nineit \def\it{\fam\itfam\nineit}%
    \textfont\slfam=\ninesl \def\sl{\fam\slfam\ninesl}%
    \textfont\bffam=\ninebf \scriptfont\bffam=\sixbf
                            \scriptscriptfont\bffam=\fivebf
                            \def\bf{\fam\bffam\ninebf}%
    \textfont\ssfam=\niness \scriptfont\ssfam=\sixss
                            \scriptscriptfont\ssfam=\sixss
                            \def\ss{\fam\ssfam\niness}%
    \normalbaselineskip=12pt
    \setbox\strutbox=\hbox{\vrule height8.0pt depth3.0pt width0pt}%
    \let\big=\ninebig
    \normalbaselines\rm}

\def\eightpoint{\def\rm{\fam0\eightrm}%
    \textfont0=\eightrm      \scriptfont0=\sixrm
                             \scriptscriptfont0=\fiverm
    \textfont1=\eighti       \scriptfont1=\sixi
                             \scriptscriptfont1=\fivei
    \textfont2=\eightsy      \scriptfont2=\sixsy
                             \scriptscriptfont2=\fivesy
    \textfont3=\tenex        \scriptfont3=\tenex
                             \scriptscriptfont3=\tenex
    \textfont\itfam=\eightit \def\it{\fam\itfam\eightit}%
    \textfont\slfam=\eightsl \def\sl{\fam\slfam\eightsl}%
    \textfont\bffam=\eightbf \scriptfont\bffam=\sixbf
                             \scriptscriptfont\bffam=\fivebf
                             \def\bf{\fam\bffam\eightbf}%
    \textfont\ssfam=\eightss \scriptfont\ssfam=\sixss
                             \scriptscriptfont\ssfam=\sixss
                             \def\ss{\fam\ssfam\eightss}%
    \normalbaselineskip=10pt
    \setbox\strutbox=\hbox{\vrule height7.0pt depth2.0pt width0pt}%
    \let\big=\eightbig
    \normalbaselines\rm}

\def\tenbig#1{{\hbox{$\left#1\vbox to8.5pt{}\right.\n@space$}}}
\def\ninebig#1{{\hbox{$\textfont0=\tenrm\textfont2=\tensy
                       \left#1\vbox to7.25pt{}\right.\n@space$}}}
\def\eightbig#1{{\hbox{$\textfont0=\ninerm\textfont2=\ninesy
                       \left#1\vbox to6.5pt{}\right.\n@space$}}}

\font\sectionfont=cmbx10
\font\subsectionfont=cmti10

\def\figurecaptionfont{\ninepoint}
\def\tablecaptionfont{\ninepoint}
\def\footnotefont{\eightpoint}

% New count registers

\newcount\equationno
\newcount\bibitemno
\newcount\figureno
\newcount\tableno

\equationno=0
\bibitemno=0
\figureno=0
\tableno=0
%\advance\pageno by -1

% Footline

\footline={\ifnum\pageno=0{\hfil}\else
{\hss\rm\the\pageno\hss}\fi}

% Section macro

\def\section #1. #2 \par
{\vskip0pt plus .20\vsize\penalty-100 \vskip0pt plus-.20\vsize
\vskip 1.6 true cm plus 0.2 true cm minus 0.2 true cm
\global\def\equationlabel{#1}
\global\equationno=0
\leftline{\sectionfont #1. #2}\par
\immediate\write\terminal{Section #1. #2}
\vskip 0.7 true cm plus 0.1 true cm minus 0.1 true cm
\noindent}

% Subsection macro

\def\subsection #1 \par
{\vskip0pt plus 0.8 true cm\penalty-50 \vskip0pt plus-0.8 true cm
\vskip2.5ex plus 0.1ex minus 0.1ex
\leftline{\subsectionfont #1}\par
\immediate\write\terminal{Subsection #1}
\vskip1.0ex plus 0.1ex minus 0.1ex
\noindent}

% Appendix macro

\def\appendix #1 \par
{\vskip0pt plus .20\vsize\penalty-100 \vskip0pt plus-.20\vsize
\vskip 1.6 true cm plus 0.2 true cm minus 0.2 true cm
\global\def\equationlabel{\hbox{\rm#1}}
\global\equationno=0
\leftline{\sectionfont Appendix #1}\par
\immediate\write\terminal{Appendix #1}
\vskip 0.7 true cm plus 0.1 true cm minus 0.1 true cm
\noindent}

% Displayed equations

\def\equation#1{$$\displaylines{\qquad #1}$$}
\def\enum{\global\advance\equationno by 1
\hfill\llap{(\equationlabel.\the\equationno)}}
\def\noenum{\hfill}
\def\next#1{\cr\noalign{\vskip#1}\qquad}

% Bibliography macro, references

\def\ifundefined#1{\expandafter\ifx\csname#1\endcsname\relax}

\def\ref#1{\ifundefined{#1}?\immediate\write\terminal{unknown reference
on page \the\pageno}\else\csname#1\endcsname\fi}

\newwrite\terminal
\newwrite\bibitemlist

\def\bibitem#1#2\par{\global\advance\bibitemno by 1
\immediate\write\bibitemlist{\string\def
\expandafter\string\csname#1\endcsname
{\the\bibitemno}}
\item{[\the\bibitemno]}#2\par}

\def\beginbibliography{
\vskip0pt plus .15\vsize\penalty-100 \vskip0pt plus-.15\vsize
\vskip 1.2 true cm plus 0.2 true cm minus 0.2 true cm
\leftline{\sectionfont References}\par
\immediate\write\terminal{References}
\immediate\openout\bibitemlist=biblist
\frenchspacing\parindent=1.8em
\vskip 0.5 true cm plus 0.1 true cm minus 0.1 true cm}

\def\endbibliography{
\immediate\closeout\bibitemlist
\nonfrenchspacing\parindent=1.0em}

% Figure and table captions

\def\figurecaption#1{\global\advance\figureno by 1
\narrower\figurecaptionfont
Fig.~\the\figureno. #1}

\def\tablecaption#1{\global\advance\tableno by 1
\vbox to 0.5 true cm { }
\centerline{\tablecaptionfont%
Table~\the\tableno. #1}
\vskip-0.4 true cm}

\def\thicktablerule{\hrule height1pt}
\def\thintablerule{\hrule height0.4pt}

\tenpoint

\immediate\openin\bibitemlist=biblist
\ifeof\bibitemlist\immediate\closein\bibitemlist
\else\immediate\closein\bibitemlist
\input biblist \fi

% current year and month

\def\thismonth{\ifcase\month\or
January\or February\or March\or April\or May\or June\or
July\or August\or September\or October\or November\or December\fi}

% Definitions and abbreviations

% Roman letters in math formulae

\def\rmd{{\rm d}}

\def\rme{{\rm e}}
\def\rmO{{\rm O}}

% Real and integer numbers

\def\gz{{\Bbb Z}}

\def\Re{{\rm Re}\,}

% Special relations and symbols

\def\proof{\noindent{\sl Proof:}\kern0.6em}

\def\frac#1#2{\hbox{$#1\over#2$}}
\def\dual{\mathstrut^*\kern-0.1em}

\def\lvec#1{\setbox0=\hbox{$#1$}
    \setbox1=\hbox{$\scriptstyle\leftarrow$}
    #1\kern-\wd0\smash{
    \raise\ht0\hbox{$\raise1pt\hbox{$\scriptstyle\leftarrow$}$}}
    \kern-\wd1\kern\wd0}
\def\rvec#1{\setbox0=\hbox{$#1$}
    \setbox1=\hbox{$\scriptstyle\rightarrow$}
    #1\kern-\wd0\smash{
    \raise\ht0\hbox{$\raise1pt\hbox{$\scriptstyle\rightarrow$}$}}
    \kern-\wd1\kern\wd0}

% Lattice derivatives

\def\nabstar#1{\nabla\kern-0.5pt\smash{\raise 4.5pt\hbox{$\ast$}}
               \kern-4.5pt_{#1}}

\def\drvstar#1{\partial\kern-0.5pt\smash{\raise 4.5pt\hbox{$\ast$}}
               \kern-5.0pt_{#1}}

\def\ldrvstar#1{\lvec{\,\partial}\kern-0.5pt\smash{\raise 4.5pt\hbox{$\ast$}}
               \kern-5.0pt_{#1}}

% Units

\def\fm{{\rm fm}}

% Constants

% Fields

% Dirac matrices

\def\diracstar#1#2{
    \setbox0=\hbox{$\gamma$}\setbox1=\hbox{$\gamma_{#1}$}
    \gamma_{#1}\kern-\wd1\kern\wd0
    \smash{\raise4.5pt\hbox{$\scriptstyle#2$}}}

% Gauge group

% Masses etc

\def\mpi{m_{\pi}}
\def\mK{m_{K}}
\def\Kbar{\kern2pt\overline{\kern-2pt K}\kern0pt}
\def\Zpi{Z_{\pi}}

\def\kpi{k_{\pi}}
\def\Kn{\Omega_n}
\def\On{{\cal O}_n}

% Interaction lagrangian

\def\Lint{{\cal L}_{\rm int}}
\def\Lw{{\cal L}_{\rm w}}
\def\Hw{H_{\rm w}}
\def\M{M}
\rightline{CERN-TH/2000-091}
\rightline{CPT-2000/PE.3984}
\rightline{LAPTH-788/00}

\vskip 0.8 true cm 
\centerline
{\Bigrm Weak transition matrix elements} 
\vskip 1.4ex
\centerline
{\Bigrm from finite-volume correlation functions}
\vskip 0.6 true cm
\centerline{\bigrm Laurent Lellouch\kern1pt$^{\rm a}$\kern1pt%
\footnote{{\footnotefont$^{\dag}$}}{\footnotefont% 
On leave from Centre de Physique Th\'eorique, CNRS Luminy,
F-13288 Marseille Cedex 9, France}
and Martin L\"uscher\kern1pt$^{\rm b}$\kern1pt%
\footnote{{\footnotefont$^{\ddag}$}}{\footnotefont% 
On leave from Deutsches Elektronen-Synchrotron DESY, 
D-22603 Hamburg, Germany}}
\vskip3ex
\centerline{$^{\rm a}$\it\kern-0pt{}LAPTH, Chemin de Bellevue, B.P.~110}
\centerline{\it F-74941 Annecy-Le-Vieux Cedex, France}
\vskip3ex
\centerline{$^{\rm b}$\it\kern-0pt{}CERN, Theory Division} 
\centerline{\it CH-1211 Geneva 23, Switzerland}
\vskip 0.8 true cm
\thintablerule
\vskip 2.0ex
\ninepoint
\leftline{\bf Abstract}
\vskip 1.0ex\noindent
The two-body decay rate of a weakly decaying particle
(such as the kaon)
is shown to be proportional to the square of a well-defined 
transition matrix element in finite volume.
Contrary to the physical amplitude, the latter
can be extracted from finite-volume correlation functions in euclidean space 
without analytic continuation.
The $K\to\pi\pi$ transitions
and other non-leptonic decays thus become accessible 
to established numerical techniques in lattice QCD.

\vskip 2.0ex
\thintablerule
\vskip -3.0ex
\tenpoint

\section 1. Introduction

The computation of the non-leptonic kaon decay rates
from first principles, using lattice QCD and numerical simulations, 
meets a number of technical difficulties
(see ref.~[\ref{DawsonEtAl}], for example).
Apart from the operator renormalization, which must be controlled
at the non-perturbative level, the central problem is 
that the computational framework is limited to 
correlation functions in euclidean space
and that there is apparently no simple relation between
the behaviour of these functions at large time separations and
the desired transition matrix elements
[\ref{MaianiTesta},\ref{CiuchiniEtAl}].

This statement
(which is often referred to as the {\it Maiani--Testa no-go theorem})
applies to very large or infinite lattices,
where the spectrum of final states is continuous.
One might think that having a
finite volume (as is unavoidable 
when numerical si\-mu\-lations are employed)
makes it even more difficult to 
extract the transition ampli\-tudes.
In the present paper we wish to show
that this is actually not so.
The key observation is that the two-pion energy spectrum 
is far from being continuous
when the lattice is only a few fermis wide.
Under these conditions, a kaon at rest cannot decay into two pions
unless one of these energy levels happens to be close
to its mass.
This is the case for certain lattice sizes,
and a simple formula then
relates the square of the corresponding 
transition amplitude in finite volume
to the physical decay rate in infinite volume.

The problem is thus reduced to calculating 
the required finite-volume transition amplitudes.
Since the initial and final states are 
isolated energy eigenstates,
these matrix elements can in principle be 
computed using established techniques, such as those
commonly employed to determine form factors.
An additional difficulty is that the
relevant two-pion states are not the lowest ones in the specified sector.
Two-particle states in finite volume have, however, previously been studied
[\ref{MontvayWeisz}--\ref{GutsfeldEtAl}]
and practical methods have been devised to calculate the higher levels.

To keep the presentation as transparent as possible,
we shall consider a simplified generic theory 
with two kinds of spinless particles, 
referred to as the kaon and the pion.
Details are given in the next section, and
we then first discuss the form of the two-pion
energy spectrum in finite volume.
This is essentially a summary of the relevant results of
refs.~[\ref{StableStates}--\ref{TwoParticleStates}].
In sect.~4 we define the transition amplitudes in finite volume 
and state the formula that relates them 
to the corresponding decay rates in infinite volume.
The following sections contain the proof of this relation and 
a discussion of its application to the physical kaon decays.

\section 2. Preliminaries

As announced above, we consider a generic
situation where there are two particles, 
the ``kaon" and the ``pion",
with spin zero and masses such that
\equation{
  2\mpi<\mK<4\mpi.
  \enum
}
We assume that 
the symmetries of the theory 
are such that the kaon is stable 
in the absence of the weak interactions
and that the pions scatter purely elastically 
below the four-pion threshold.
The weak interactions, described by a local effective
lagrangian $\Lw(x)$, then allow the kaon to decay
into two pions.
The corresponding transition amplitude is 
\equation{
  T(K\to\pi\pi)=
  \langle \pi\,p_1,\pi\,p_2\,\hbox{out}|
  \Lw(0)|K\,p\rangle,
  \enum
}
with $p_1$, $p_2$ and $p$ the four-momenta of the pions and the
kaon. We shall only be interested in the physical case where the total momentum
$p=p_1+p_2$ is conserved. Lorentz invariance and the kinematical constraints
then imply that the transition amplitude
is independent of the momentum configuration.

The meson states in eq.~(2.2) are normalized according to
the standard relativistic conventions (appendix A)
and their phases are constrained by the LSZ formalism.
In the case of the pions, for example,
one assumes that there exists an interpolating 
hermitian field $\varphi(x)$ such that
\equation{
  \langle 0|\varphi(x)|\pi\,p\rangle=\sqrt{\Zpi}\kern1pt\rme^{-ipx}
  \enum
}
for some positive constant $\Zpi$. 
If the phase of the kaon states is chosen in the same way,
the CPT symmetry implies
\equation{
    T(K\to\pi\pi)=A\rme^{i\delta_0}
    \enum
}
with $A$ real and $\delta_0$ the $S$-wave 
scattering phase shift of the outgoing pion state.
The decay rate is then given by the usual expression
\equation{
  \Gamma={\kpi\over16\pi\mK^2}\left|A\right|^2, 
  \qquad
  \kpi\equiv\frac{1}{2}\sqrt{\mK^2-4\mpi^2},
  \enum
}
proportional to the pion momentum $\kpi$ in the centre-of-mass frame.

\section 3. Two-pion states in finite volume

In a spatial box of size $L\times L\times L$ with 
periodic boundary conditions, the eigenvalues of the 
total momentum operator are integer multiples of $2\pi/L$.
The energy spectrum is also discrete in this situation,
with level spacings that can be appreciable.
In the following we consider the subspace of states
with zero total momentum and trivial transformation behaviour
under cubic rotations and reflections.

The energy spectrum of the two-pion states in this sector
below the inelastic threshold $W=4\mpi$
has been studied in detail in 
refs.~[\ref{StableStates}--\ref{Resonances}].
In particular, for the lowest energy value the expansion
\equation{
  W=2\mpi-{4\pi a_0\over\mpi L^3}
  \left\{1+c_1{a_0\over L}
          +c_2{a_0^2\over L^2}\right\}+\rmO(L^{-6}),
  \enum
  \next{2.5ex}
  c_1=-2.837297,
  \qquad
  c_2=6.375183,
  \enum
}
has been obtained, where 
\equation{
  a_0=\lim_{k\to0}{\delta_0(k)\over k}
  \enum
}
is the $S$-wave scattering length
(here and below the scattering phase is considered to be a function 
of the pion momentum $k$ in the centre-of-mass frame).
The higher energy values in the elastic region are determined through
\equation{
  W=2\sqrt{\mpi^2+k^2},
  \enum
  \next{2.5ex}
  n\pi-\delta_0(k)=\phi(q),
  \qquad q\equiv{kL\over2\pi},
  \enum 
}
where $n=1,2,\ldots$ labels the energy levels in increasing order
and the angle $\phi(q)$ is a known kinematical function
(appendix B).
Apart from the lowest level,
the energy spectrum at any given value of $L$ is thus obtained
by inserting the solutions $k$ of eq.~(3.5) 
in eq.~(3.4)\kern1pt%
\footnote{$\dag$}{\footnotefont%
Similar formulae have been derived for the spectrum 
in the subspaces of states with non-zero total momentum
[\ref{RummukainenGottlieb}]. The extension of our results to 
these sectors could 
give further insight into the connection between finite and 
infinite volume matrix elements and may prove useful in practice.}.

All these results are valid up to terms that vanish exponentially
at large $L$. Box sizes a few times larger than
the diameter of the pion should be safe from these corrections.
Equation (3.5) moreover assumes that the scattering phases $\delta_l$ for 
angular momenta $l\geq4$ are small in the elastic region, 
which is usually the case
since $\delta_l$ is proportional to $k^{2l+1}$ at low energies.

\topinsert
\vbox{
\vskip-0.5 true cm
 
\centerline{
\epsfxsize=8.0 true cm
\epsfbox{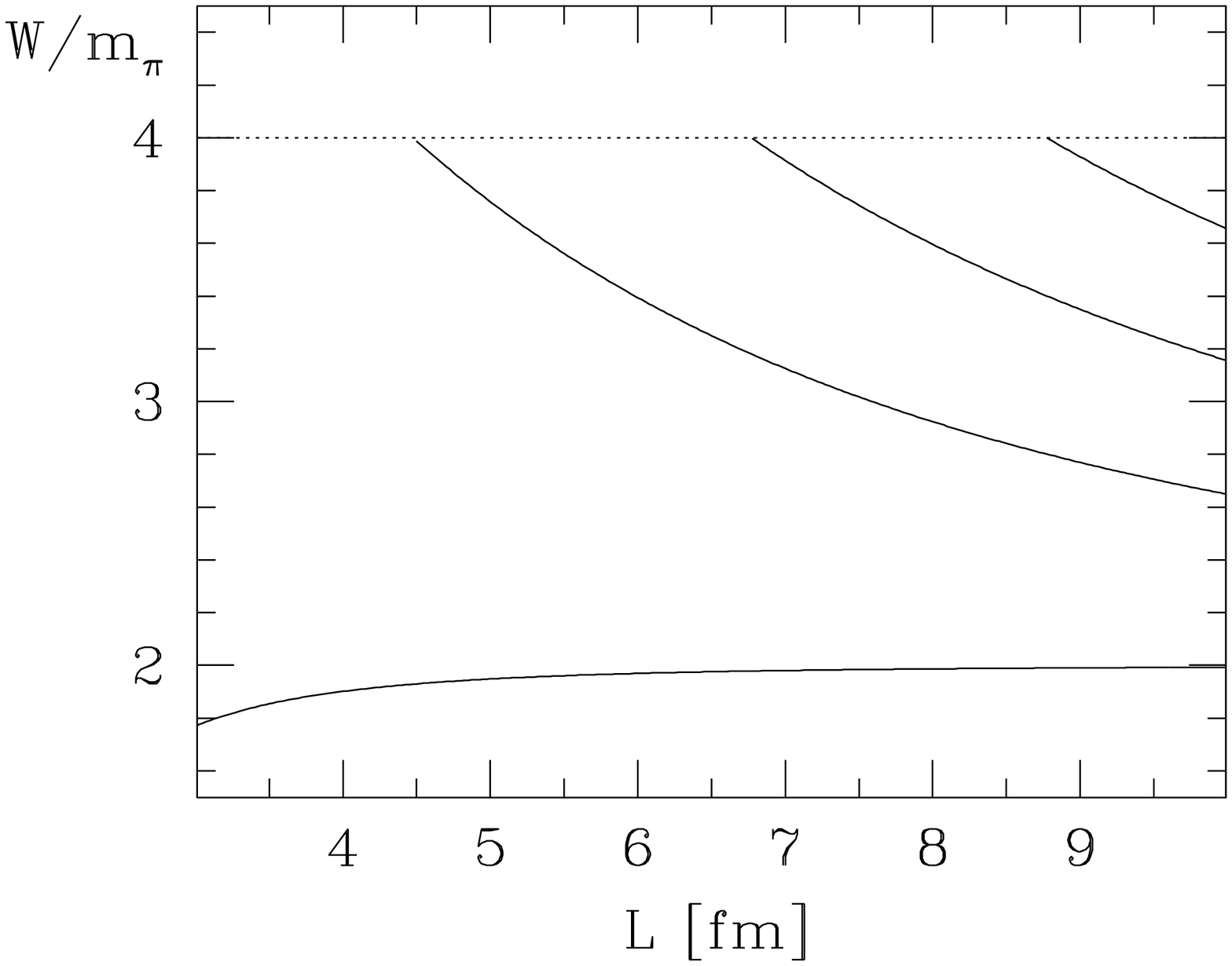}
}
\vskip-0.8 true cm
\figurecaption{%
Two-pion energy spectrum in QCD below the inelastic threshold,
in the sector with isospin $0$,
calculated from eqs.~(3.1)--(3.5) with the scattering phase shift
given by next-to-leading order chiral perturbation theory.
The levels shown in this plot are all non-degenerate.
}
\vskip0.3 true cm
}
\endinsert

For illustration, let us consider QCD with 
three flavours of quarks, unbroken isospin symmetry
and quark masses such that the masses of the charged pions and kaons
coincide with their physical values.
In the subspace with isospin $0$, the two-pion energy spectrum 
is then given by eqs.~(3.1)--(3.5),
with $\delta_0$ the appropriate pion scattering phase.
If we insert the phase shift that is obtained at one-loop order of
chiral perturbation theory [\ref{GasserLeutwyler}--\ref{KnechtEtAl}],
this yields the curves shown in fig.~1.
For any other reasonable choice of the scattering phase
the plot would look essentially the same,
because the interaction effects are proportional to $1/L^3$ and thus
tend to be small.
Note that the spacing between successive levels
is quite large. One is clearly very far away from having a continuous
spectrum when $L\leq10\;\fm$.

\section 4. Kaon decays in finite and infinite volume

Let us imagine that a state $|K\rangle$ describing a kaon in finite volume 
with zero momentum has been prepared at time $x_0=0$.
In the absence of the weak interactions, this is an energy eigenstate
(and thus a stationary state) with energy $\mK$.
However, through the interaction hamiltonian
\equation{
  \Hw=\int_{x_0=0}\rmd^3x\,\Lw(x),
  \enum
}
the time evolution of the state
becomes non-trivial and it starts to mix with
the other eigenstates
of the unperturbed hamiltonian.
It is straightforward to work this out 
using ordinary time-dependent perturbation theory.
For the transition probability at time $x_0=t$ to any finite-volume
two-pion state $|\pi\pi\rangle$ with energy $W$, the result
\equation{
  P(K\to\pi\pi)=
  4\left|\langle\pi\pi|\Hw|K\rangle\right|^2
  {\sin^2\bigl({1\over2}\omega t\bigr)\over\omega^2},
  \qquad \omega\equiv W-\mK,
  \enum
}
is then obtained
(in this equation the states are assumed to be normalized to unity and 
higher-order weak-interaction effects have been neglected).

From eq.~(4.2) one infers that the transition probabilities 
tend to be very small
unless the energy of one of the two-pion final states happens to be
close to the kaon mass. 
Recalling fig.~1, it is clear that 
this will be the case only for certain box sizes $L$.
In the following we 
focus on these special values of $L$ and introduce the 
associated transition matrix element
\equation{
  \M=\langle\pi\pi|\Hw|K\rangle,
  \enum
}
where {\it both states are normalized to unity}\/ as before, while 
their phase will not matter and can be chosen arbitrarily.
Since $W=\mK$ in this case, eq.~(4.2) becomes
\equation{
  P(K\to\pi\pi)=\left|\M\right|^2t^2
  \enum
}
and the kaon will thus have an appreciable probability to decay into 
the two-pion state if one waits long enough 
(the formula breaks down at very large times, 
because the higher-order terms are then no longer negligible).

The central result obtained in the present paper is 
that the finite-volume matrix element $\M$ 
is related to the decay rate of the kaon in infinite volume through
\equation{
  \left|A\right|^2=8\pi\left\{q{\partial\phi\over\partial q}+
  k{\partial\delta_0\over\partial k}\right\}_{k=\kpi}\kern-1pt
  \left({\mK\over\kpi}\right)^3\left|\M\right|^2
  \enum
}
[cf.~eqs.~(2.5),(3.5)].
The relation holds under the same premises as eq.~(3.5)
and the comments made in sect.~3 thus apply here too.
Another restriction is that the two-pion final state has to be 
non-degenerate in the specified sector
of the un\-per\-turbed theory. This condition is satisfied  
for $n<8$ [\ref{TwoParticleStates}],
but degeneracies can occur at higher level numbers 
and the formula then ceases to be valid.

In principle eq.~(4.5) allows one to compute
the kaon decay rate in infinite volume by studying 
the theory in finite volume. 
Note that in the course of such a calculation
it should also be possible to determine
the two-pion energy spectrum and thus the scattering phase $\delta_0$
in the elastic region. 

The proportionality factor in eq.~(4.5) essentially accounts for 
the different normalizations of the particle 
states in finite and infinite volume.
One can easily check this in the free theory,
where the pion self-interactions are neglected.
In this case and for $n\leq6$,
the $n$-th two-pion energy level passes through $\mK$ at
\equation{
  L={2\pi\over\kpi}\sqrt{n}.
  \enum
}
Equation (4.5) then assumes the form
\equation{
  \left|A\right|^2={4\over\nu_n}
  \left(\mK L\right)^3\left|\M\right|^2,
  \enum
  \next{2ex}
  \nu_n\equiv\hbox{number of integer vectors $\bf z$ with ${\bf z}^2=n$},
  \enum
}
which is precisely what is derived from 
the relative normalizations of the plane waves in finite and infinite volume
that describe the (non-interacting) kaon and pion states
(sect.~6).

\section 5. Proof of eq.~(4.5)

The interpretation of the proportionality factor in eq.~(4.5) given
above also applies in the interacting case.
This follows from the fact that the transition matrix elements
probe the $S$-wave component of the two-pion wave function
near the origin and that this component is the same in finite and
infinite volume apart from its phase and normalization.
The latter can be worked out explicitly  
in the framework of refs.~[\ref{StableStates}--\ref{TwoParticleStates}],
but the calculation is rather involved and will not be 
presented here.

Instead we shall go through a different argument, 
where one studies the influence of the weak interaction
on the energy spectrum in finite volume.
This can be done directly, using ordinary
perturbation theory,
or one may start from eq.~(3.5) and take the 
weak-interaction effects on the scattering phase into account.
The combination of the results of these 
calculations then yields eq.~(4.5).

As already mentioned in sect.~2, the kaon is assumed to
carry a quantum number (alias strangeness)
that forbids its decay into pions in the unperturbed theory.
Since only the strangeness-changing part of the weak interaction 
lagrangian contributes to the kaon transition amplitudes,
all other terms may be dropped without loss.
The matrix elements of 
the weak hamiltonian $\Hw$ between states with the same
strangeness are then all equal to zero. As a consequence  
most energy values in finite volume are affected by 
the weak interaction only to second order.

First order energy shifts do occur, however, if there are degenerate states
at lowest order that mix under the action of $\Hw$.
This is the case
at the values of $L$ where one of the two-pion energy values
coincides with the kaon mass, i.e.~at the special points considered
in the preceding section.
Degenerate perturbation theory then yields
\equation{
  W=\mK\pm\left|\M\right|+\dots
  \enum
}
for the first order change of these energy values
(here and below the ellipses denote higher-order terms that 
do not contribute to the final results).

The energy shifts (5.1) can also be calculated
by including the weak corrections to the scattering phase
on the left-hand side of eq.~(3.5). From the above 
one infers that the solutions of eq.~(3.5) we are interested in  
are given by
\equation{
  k=\kpi\pm\Delta k+\ldots,
  \qquad
  \Delta k\equiv
  {\mK\over4\kpi}\left|M\right|.
  \enum
}
Compared to the kaon resonance width (which is of second order
in the weak interaction),
these values of $k$ are far away from the kaon pole.
The weak corrections to 
the pion scattering amplitude in the relevant range of energies
are hence small and can be safely computed by working out the 
perturbation expansion in powers of the interaction lagrangian. 

\topinsert
\vbox{
\vskip0.0 true cm
 
\centerline{
\epsfxsize=3.2 true cm
\epsfbox{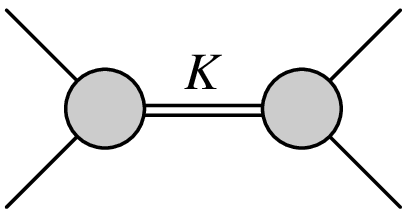}}
 
\vskip0.3 true cm
\figurecaption{%
Kaon resonance contribution to the elastic pion scattering amplitude
in the $s$-channel. The diagram appears at second order of the expansion
in powers of the weak interaction, with 
the  bubbles representing the first-order $K\pi\pi$ vertex function.
}
\vskip0.1 true cm
}
\endinsert

One might think that these corrections
are all of second or higher order, because the 
interaction is strangeness-changing. 
The reason this is not so is that the kaon propagator in a diagram
like the one shown in fig.~2 evaluates to 
\equation{
  {iZ_K\over p^2-\mK^2}
  =\pm{iZ_K\over2\mK\left|\M\right|}+\ldots
  \enum
}
at the energies (5.1) and thus reduces the effective order
of the term by $1$. 
This diagram is in fact the only one that yields a
first-order contribution to the scattering amplitude.
It can be calculated by noting that the momenta
flowing into the three-point vertices are all on shell 
up to higher-order corrections. The vertices 
are hence proportional to the kaon decay amplitude $A$.
Together with eq.~(5.3) this leads to the result
\equation{
  \bar{\delta}_0(k)=
  \delta_0(k)\mp{\kpi\left|A\right|^2\over
  32\pi\mK^2\left|\M\right|}+\ldots
  \quad(\hbox{mod}\,\pi)
  \enum
}
for the scattering phase in the full theory at the point (5.2)
(as in the previous section, 
$\delta_0$ stands for the phase shift in the 
unperturbed theory).

We now replace $\delta_0$ 
in eq.~(3.5) by $\bar{\delta}_0$ and expand all terms
in powers of the weak interaction. 
The lowest-order terms cancel while,
at first order, the equation implies
\equation{
  -\Delta k\left\{{\partial\delta_0(k)\over
  \partial k}\right\}_{k=\kpi}
  +{\kpi\left|A\right|^2\over
  16\pi\mK^2\left|\M\right|}
  =\Delta k\left\{\partial\phi(q)\over\partial k\right\}_{k=\kpi}.
  \enum
}
This is easily seen to be equivalent to eq.~(4.5) after substituting
the expression (5.2) for $\Delta k$ and
we have thus proved this relation.

\section 6. Verification of eq.~(4.5) in perturbation theory

In a low-energy effective theory, such as the chiral non-linear
$\sigma$-model, it is possible to obtain an independent check on
eq.~(4.5) by working out the transition amplitudes in finite and
infinite volume in perturbation theory.
Since this calculation does not rely on 
any of the results presented above,
it can provide additional confidence in the correctness of the equation.
Perturbation theory may also prove helpful when considering 
more complicated situations, 
where one has several decay channels or particles
with non-zero spin.
In this section we describe how such a calculation 
goes, without giving too many details.

\subsection 6.1 Specification of the model

The two-pion energy spectrum and the 
proportionality factor in eq.~(4.5) depend on the final-state
interactions only through the phase shift $\delta_0$.
All other properties of the pion interactions do not matter
and to check the equation we may thus
consider an arbitrary effective meson theory with the correct
particle spectrum.

For the pion interaction lagrangian the simplest choice is
\equation{
  \Lint(x)=\frac{1}{4!}\lambda\varphi(x)^4,
  \enum
}
where $\varphi(x)$ denotes the pion field and 
$\lambda$ the bare coupling. 
To make the perturbation expansion completely well-defined,
we introduce a Pauli--Villars cutoff $\Lambda$.
At tree level the euclidean pion propagator
is then given by
\equation{
  \int\rmd^4x\,\rme^{-ipx}\langle\varphi(x)\varphi(0)\rangle=
  {1\over m^2+p^2}-{1\over\Lambda^2+p^2},
  \enum
}
with $m$ the bare mass of the pion (its physical mass is denoted by $\mpi$
as before). The cutoff should be large enough
so that ghost particles cannot be produced at energies below the 
four-pion threshold, but in view of the universality of 
eq.~(4.5) there is no need to take $\Lambda$ to infinity at the 
end of the calculation. 

As far as the kaon is concerned, the least complicated possibility is
to describe it by a hermitian free field $\theta(x)$ with mass $\mK$
and to take
\equation{
  \Lw(x)=\frac{1}{2}g\theta(x)\varphi(x)^2
  \enum
}
as the weak-interaction lagrangian. One then first has to expand 
the transition am\-pli\-tude (2.2) in powers of $\lambda$, but
we shall not discuss this here 
since the calculation is completely standard.
The way to 
obtain the perturbation expansion of
the finite-volume matrix element (4.3) 
may be less obvious, however,
and we thus proceed to explain this in some detail.

\subsection 6.2 Two-pion states

In finite volume the low-lying two-pion energy eigenstates
with zero total momentum and trivial transformation 
behaviour under the cubic group may be labelled by 
an integer $n=0,1,2\ldots$ such that the associated energies $W_n$ 
increase monotonically with $n$. 
We denote these states by $|\pi\pi\,n\rangle$ and assume that 
they have unit norm.

To lowest order in $\lambda$,
the energy values are determined through the free
energy-momentum relation and the relative momentum of the pions.
Since we only consider cubically invariant states,
any two momenta that are related to each other 
by a cubic transformation describe the same state.
For $n\leq6$ the momenta in the set
\equation{
 \Kn=\left\{{\bf k}=2\pi{\bf z}/L\bigm| 
 {\bf z}\in\gz^3, {\bf z}^2=n\right\}
 \enum
}
are all equivalent in this sense. The corresponding state
is thus non-degenerate and one concludes from this that 
\equation{
  W_n=2\sqrt{m^2+n(2\pi/L)^2}+\rmO(\lambda),
  \qquad 0\leq n\leq6.
  \enum
}
In the following, our attention will be restricted to these levels.

\topinsert
\vbox{
\vskip0.0 true cm
 
\centerline{
\epsfxsize=9.0 true cm
\epsfbox{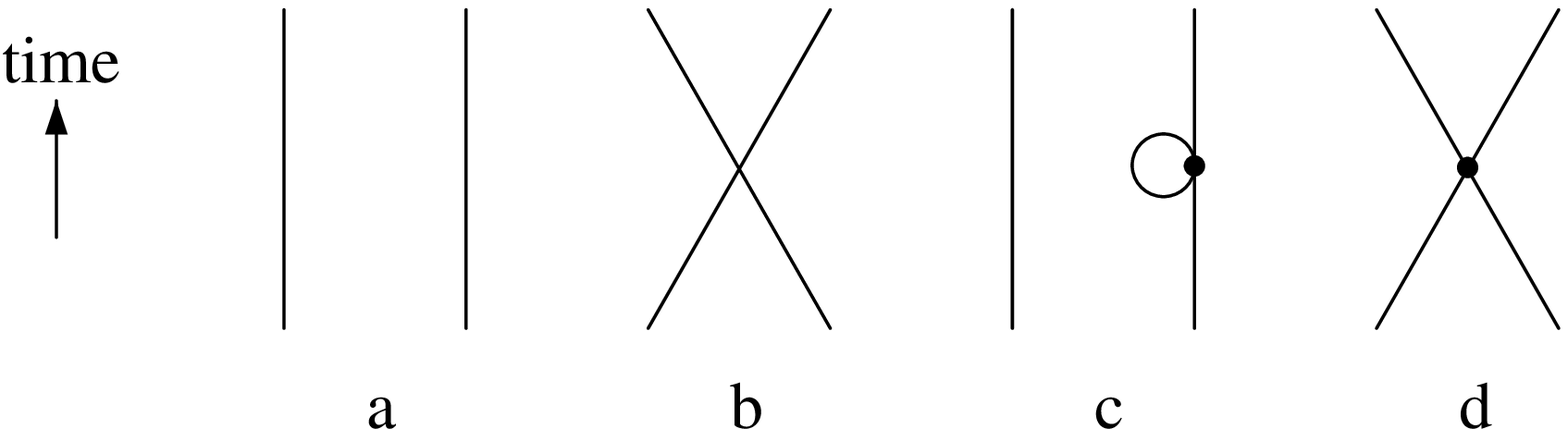}}
 
\vskip0.3 true cm
\figurecaption{%
Feynman diagrams contributing to the correlation function (6.7).
The lines represent the free pion propagator in the time-momentum 
representation (6.8) and the filled circles the
self-interaction vertex. All external lines end at times $x_0$
or $y_0$.
}
\vskip0.0 true cm
}
\endinsert

The corresponding energy eigenstates 
$|\pi\pi\,n\rangle$ can be created from the vacuum 
by applying the operators
\equation{
  \On(x_0)=\sum_{{\bf k}\in\Kn}
  \int_0^L\rmd^3x\kern1pt\rmd^3y\,
  \rme^{i{\bf k}({\bf x-y})}\varphi(x_0,{\bf x})\varphi(x_0,{\bf y}).
  \enum
}
Note that $\On(x_0)$ couples
to all two-pion states in the given sector, since there
are no quantum numbers that would forbid this.
In euclidean space and at large time separations $x_0-y_0$,
its connected two-point function is thus given by
\equation{
  \langle\On(x_0)\On(y_0)\rangle_{\rm con}=
  \sum_{l=0}^6
  \left|\langle 0|\On(0)|\pi\pi\,l\rangle\right|^2
  \rme^{-W_{l}(x_0-y_0)}+\ldots,
  \enum
}
where the ellipses stand for more rapidly decaying terms.

The perturbation expansion of 
the two-pion energy $W_n$ and the associated matrix element 
$\left|\langle 0|\On(0)|\pi\pi\,n\rangle\right|$
may now be obtained by expanding the left-hand side of eq.~(6.7)
in Feynman diagrams in the standard way.
If one uses the time-momentum representation
\equation{
  \int_0^L\rmd^3 x\,
  \rme^{-i{\bf px}}\langle\varphi(x)\varphi(0)\rangle=
  {\rme^{-\omega_{\bf p}|x_0|}\over2\omega_{\bf p}}
  -(m\leftrightarrow\Lambda),
  \qquad
  \omega_{\bf p}\equiv\sqrt{m^2+{\bf p}^2},
  \enum
}
for the tree-level pion propagator,
the diagrams evaluate to a sum of exponentials.
The desired expansions can then be read off from the 
coefficients of the exponential factor that corresponds to the 
$n$-th level.

To leading order the diagrams a and b in fig.~3
yield the expected expression (6.5)
for the two-pion energy and 
\equation{
  \left|\langle 0|\On(0)|\pi\pi\,n\rangle\right|=
  \sqrt{2\nu_n}L^3/W_n+\rmO(\lambda)
  \enum
}
for the matrix element [cf.~eq.~(4.8)].
At the next order in the coupling, there are two types of diagrams. 
Diagram c and three further diagrams of this kind
amount to an additive renormalization
of the pion mass by a term that is independent of $L$
up to exponentially small corrections [\ref{StableStates}].
Such contributions are neglected here and the renormalization
is thus equivalent to replacing
$m$ by $\mpi$ in the tree-level expressions.
One is then left with the diagram d,
which can be worked out analytically in a few lines.

\topinsert
\vbox{
\vskip0.0 true cm
 
\centerline{
\epsfxsize=8.0 true cm
\epsfbox{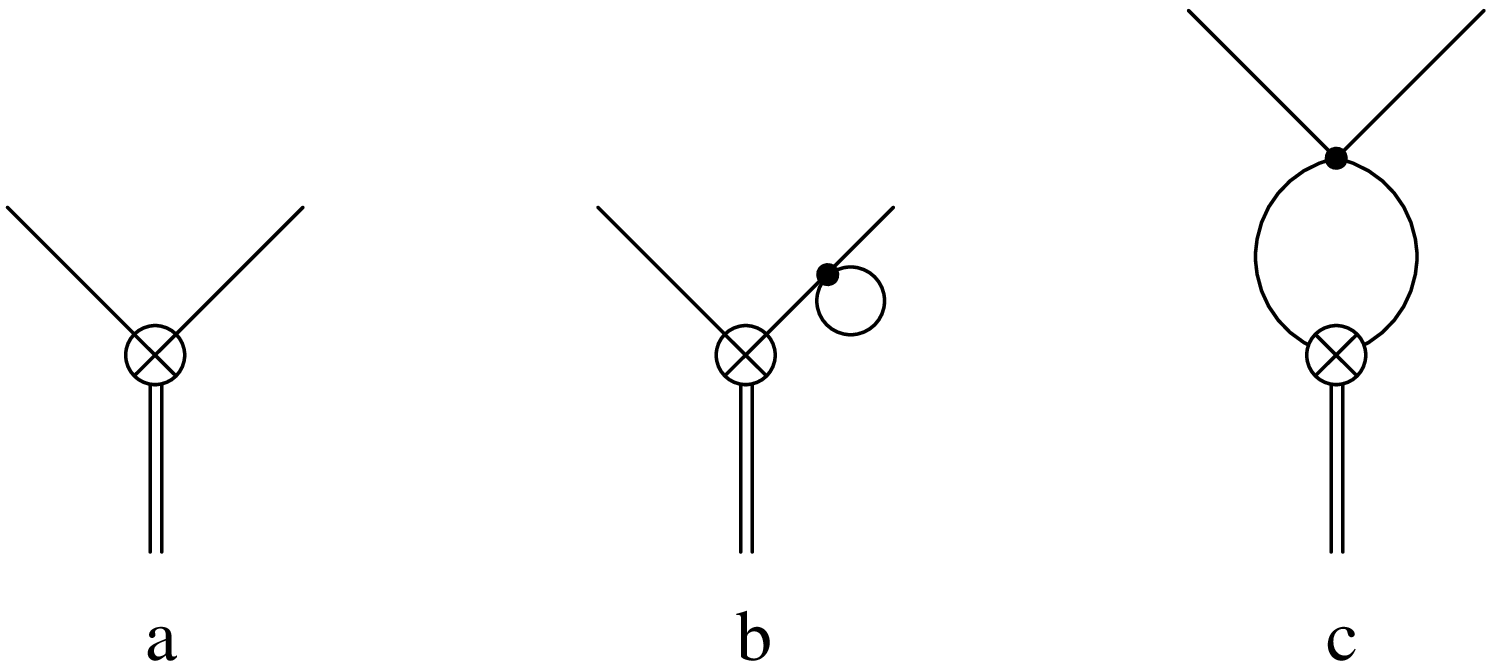}}
 
\vskip0.3 true cm
\figurecaption{%
Diagrams contributing to the correlation function (6.10). 
The double line represents the 
kaon propagator and the circled cross the weak interaction vertex
at the origin. All other graphical elements are as in fig.~3.
}
\vskip0.0 true cm
}
\endinsert

\subsection 6.3 Transition matrix element

The finite-volume transition matrix element (4.3) can be 
computed by studying the euclidean correlation function
\equation{
  \int_0^L\rmd^3y\,\langle\On(x_0)\Lw(0)\theta(y)\rangle_{\rm con}=
  \noenum
  \next{2ex}
  \qquad\sum_{l=0}^6\rme^{-W_lx_0+\mK y_0}
  \langle0|\On(0)|\pi\pi\,l\rangle\langle\pi\pi\,l|\Hw|K\rangle
  \langle K|\theta(0)|0\rangle+\ldots
  \enum
}
at large $x_0$ and large negative $y_0$. 
As in the case of the two-pion states, the terms we are
interested in are found by looking for the 
appropriate exponential factor.

To lowest order diagram a in fig.~4 yields
\equation{
  |\langle\pi\pi\,n|\Hw|K\rangle|=
  {g\sqrt{\nu_n}\over2W_n\sqrt{\mK L^3}}\left\{1+\rmO(\lambda)\right\}.
  \enum
}
The pion mass in this expression is renormalized by 
the tadpole insertions at the next order (diagram b and its mirror image).
Diagram c, the only other diagram at this order,
may be evaluated by inserting the time-momentum representation 
for the external and also the internal lines.
Apart from various simple terms, one then ends up with 
the momentum sum
\equation{
  S_n=L^{-3}\sum_{{\bf p}\notin\Kn}
  \left\{{1\over\omega_{\bf p}({\bf p}^2-{\bf k}^2)}
  -R_{\Lambda}(\hbox{${\bf p}^2,{\bf k}^2$})\right\},\qquad {\bf k}\in\Kn,
  \enum
}
where $R_{\Lambda}$ is an expression that
arises from the Pauli-Villars regularization. 

A general summation formula proved in ref.~[\ref{ScatteringStates}]
allows one to compute such sums up to terms that vanish
more rapidly than any power of $1/L$.
The precise form of $R_{\Lambda}$ is not important for this.
One only needs to know that it is a smooth function of $\bf p$ and $\bf k$
and that it makes the sum absolutely convergent.
The result
\equation{
  S_n=\int{\rmd^3p\over(2\pi)^3}
  \left\{{1\over2\omega_{\bf p}}
  \left[{1\over{\bf p}^2-{\bf k}^2+i\epsilon}
       +{1\over{\bf p}^2-{\bf k}^2-i\epsilon}\right]
  -R_{\Lambda}(\hbox{${\bf p}^2,{\bf k}^2$})\right\}
  \noenum
  \next{2.5ex}
  \kern2.5em
  +{z_n\over4\pi^2\omega_{\bf k}L}
  +{\nu_n\over2(\omega_{\bf k}L)^3}
  +{\nu_n\over L^3}R_{\Lambda}(\hbox{${\bf k}^2,{\bf k}^2$})
  \enum
}
is then obtained, with the constant $z_n$ given by
\equation{
  z_n=\lim_{q^2\to n}\left\{
  \sqrt{4\pi}{\cal Z}_{00}(1;q^2)+{\nu_n\over q^2-n}\right\}
  \enum
}
(the zeta function ${\cal Z}_{00}(s;q^2)$ is defined in appendix B).

\subsection 6.4 Final steps

To check eq.~(4.5) one has to tune the box size
so that $W_n=\mK$ for a specified level number $n$.
This condition determines $L$ order by order in the coupling.
The perturbation expansion of the right-hand side of eq.~(4.5)
is then obtained by 
inserting this series in the proportionality factor and
the perturbative expressions 
for the matrix element $|\langle\pi\pi\,n|\Hw|K\rangle|$.

To lowest order, the box size is given by eq.~(4.6) and 
the function $\phi'(q)$ in the proportionality
factor is thus to be expanded around $q=\sqrt{n}$.
This generates a term proportional to $z_n$, which cancels  
the corresponding term in eq.~(6.13).
The integral in this equation matches with
the contribution to the transition amplitude $A$
of the infinite-volume diagram with the topology of diagram c.
All other terms that occur at first order in the coupling 
cancel and one finds that eq.~(4.5) holds as expected.

\section 7. Application to the physical kaon decays

Compared to the generic theory considered so far, 
the situation in the case of the physical kaon decays
is complicated by the fact that 
there are several decay channels.
To a first approximation we may however assume that 
isospin is an exact symmetry in the absence of the weak interactions.
The decay channels can then be separated from each other
by passing to a basis of states with definite quantum numbers.

As an example we discuss the 
CP-conserving decays of the neutral kaon into
two-pion states with isospin $0$ and $2$.
The corresponding decay amplitudes, $A_0$ and $A_2$, are related
to the physical transition matrix elements through
\equation{
  T(K^0_S\to\pi^{+}\pi^{-})=
  {2\over\sqrt{6}}\kern1pt A_0\rme^{i\delta_0^0}
  +{1\over\sqrt{3}}\kern1pt A_2\rme^{i\delta_0^2},
  \enum
  \next{2.5ex}
  T(K^0_S\to\pi^{0}\pi^{0})=
  -{2\over\sqrt{6}}\kern1pt A_0\rme^{i\delta_0^0}
  +{2\over\sqrt{3}}\kern1pt A_2\rme^{i\delta_0^2}.
  \enum
}
In these equations
$\delta_0^I$ denotes the $S$-wave pion scattering phase in the channel
with isospin $I$ and the normalization and phase
conventions are as in sect.~2.

In the sector of two-pion states
with isospin $I$, zero electric charge,  
zero total momentum and trivial transformation behaviour under
cubic rotations and reflections,
the energy spectrum in finite volume is determined by the equations
that we have previously discussed, with
$\delta_0$ replaced by $\delta^I_0$.
At the points where one of these energy levels 
passes through $\mK$, 
we define the associated transition matrix element
\equation{
  M_I=\langle(\pi\pi)_I|\Hw|K^0\rangle,
  \enum
}
where it is understood that the states are normalized to unity
and that $\Hw$ is the CP-conserving part of the 
effective weak hamiltonian.
With these conventions, the physical amplitudes are given by
\equation{
  \left|A_I\right|^2=8\pi\left\{q{\partial\phi\over\partial q}+
  k{\partial\delta_0^I\over\partial k}\right\}_{k=\kpi}\kern-1pt
  \left({\mK\over\kpi}\right)^3\left|\M_I\right|^2.
  \enum
}
Note that $A_0$ and $A_2$ are real and only their relative sign
is observable. Up to this sign, the complete information 
can thus be retrieved from the matrix elements and 
the energy spectrum in finite volume.

\topinsert
%Blanke Zahl
\newdimen\digitwidth
\setbox0=\hbox{\rm 0}
\digitwidth=\wd0
\catcode`@=\active
\def@{\kern\digitwidth}
\tablecaption{%
Calculation of the proportionality factor in eq.~(7.4) 
at the first level crossing
}
\vskip2ex
$$\vbox{\settabs\+x&xxxxx&%
                  xx&xxxxx&%
                  xxxx&xxxxx&%
                  xxx&xxxxxxxxxxxx&%
                  x&xxxxxxxxxxxx&%
                  x\cr
\thicktablerule
\vskip1ex
                \+& \hfill $I$ \hfill
                 && \hfill $L$ [fm] \hfill
                 && \hfill $q$ \hfill
                 && \hfill $q\partial\phi/\partial q$ \hfill
                 && \hfill $k\partial\delta_0^I/\partial k$ \hfill
                 &  \cr
\vskip1.0ex
\thintablerule
\vskip1.5ex
  \+& \hfill $0$ \hfill
  &&  \hfill $5.34$ \hfill 
  &&  \hfill $0.89$ \hfill
  &&  \hfill $4.70$ \hfill
  &&  \hfill $\phantom{-}1.12$ \hfill
  &\cr
  \+& \hfill $2$ \hfill
  &&  \hfill $6.09$ \hfill 
  &&  \hfill $1.02$ \hfill
  &&  \hfill $6.93$ \hfill
  &&  \hfill $-0.09$ \hfill
  &\cr
\vskip1ex
\thicktablerule
}$$
\endinsert

For illustration,
let us suppose that the scattering phases $\delta_0^I$ are 
accurately described by the one-loop formulae of
chiral perturbation theory [\ref{GasserLeutwyler}--\ref{KnechtEtAl}].
The two-pion energy spectrum in the subspaces with isospin $I$
and the box sizes $L$, where the next-to-lowest levels in these sectors 
(the ones with level number $n=1$) 
coincide with the kaon mass, can then be calculated.
After that the proportionality factor 
in eq.~(7.4) is easily evaluated (table~1)
and one ends up with
\equation{
  |A_0|=44.9\times|\M_0|,
  \enum
  \next{2ex}
  |A_2|=48.7\times|\M_2|,
  \enum
  \next{2ex}
  |A_0/A_2|=0.92\times|M_0/M_2|.
  \enum
}
As can be seen from these figures, the large difference
between the scattering phases in the two isospin channels 
(about $45^{\circ}$ at $k=\kpi$) does not lead to 
a big variation in the proportionality factors.
In fact, if we set the scattering phases to zero altogether,
eqs.~(4.6)--(4.8) give
$|A_I|=47.7\times|M_I|$ for $n=1$, which is not far from 
the results quoted above.

The proportionality factor in eq.~(7.4) thus appears to be 
only weakly dependent on the final-state interactions.
In particular, if the theory is to reproduce the $\Delta I=1/2$
enhancement, the large factor has to come from the ratio of the 
finite-volume matrix elements $\M_I$.

\section 8. Concluding remarks

Finite-volume techniques have been used in  
lattice field theory for many years and have long proved
to be a most effective tool.
It may well be that weak transition matrix
elements are also best approached in this way.
For two-body decays a concrete proposition
along this line has been made here,
which is conceptually satisfactory and which we believe 
has a fair chance to work out in practice.

In the case of the physical kaon decays,
the proportionality factor
relating the transition matrix elements in 
finite and infinite volume turned out to be nearly the same 
in the two isospin channels. This may be surprising at first sight,
since the interactions of the pions in the isospin $0$ state 
are much stronger than in the isospin $2$ state.
One should, however, take into account the fact
that the comparison is made at box sizes $L$ greater than $5$~fm.
It is hence quite plausible that 
the finite-volume matrix elements already include 
most of the final-state interaction effects
(such as the ones recently discussed 
in refs.~[\ref{PallantePich}--\ref{BurasEtAl}]).
Apart from a purely kinematical factor,
an only small correction is then required to pass to the 
matrix elements in infinite volume.

Since the unitarity of 
the underlying field theory has been essential for our argumentation,
it is not obvious that eq.~(4.5) holds in quenched QCD.
As usual, however, one expects to be safe from the deficits of
the quenched approximation when the quark masses are not too small
and our results should then be applicable. 
An investigation of the problem in quenched chiral perturbation theory,
following refs.~[\ref{BernardGolterman},\ref{GoltermanLeung}],
may be worth while at this point to find out where precisely the 
unphysi\-cal effects set in.

As a final comment we note that
the ideas developed in this paper may 
also be applied to baryon decays, such as 
$\Lambda\to N\pi$, $\Sigma\to N\pi$ and $\Xi\to\Lambda\pi$,
as well as to any other decay where the
particles in the final state scatter only elastically.
Depending on the kinematical details,
the relation between the finite and infinite 
volume transition matrix elements
may, however, assume a slightly different form.

\appendix A

The components of four-vectors in real and euclidean space
are labelled by an index running from 0 to 3.
Bold-face types denote the spatial parts of the corresponding 
four-vectors and
scalar products are always taken with euclidean metric, except
for Lorentz vectors in real space where 
$xy=x_0y_0-{\bf x}{\bf y}$.

States $|\kern1pt p\rangle$ in infinite volume describing a 
spinless particle, with mass $m$ and four-momentum
\equation{
  p=(p_0,{\bf p}),\qquad
  p_0=\sqrt{m^2+{\bf p}^2}>0,
  \enum
}
are normalized in such a way that
\equation{
  \langle p\kern1pt|\kern1.5pt p'\rangle=2p_0(2\pi)^3
  \delta({\bf p}-{\bf p}').
  \enum
}
Particle states in finite volume are always normalized to unity.

In the centre-of-mass frame, the elastic scattering amplitude of
two spinless particles of mass $m$ may be expanded 
in partial waves according to 
\equation{
  T=16\pi W\sum_{l=0}^{\infty}(2l+1)P_l(\cos\theta)t_l(k),
  \qquad
  W=2\sqrt{m^2+k^2},
  \enum
}
where $W$ denotes the total energy of the particles,
$\theta$ the scattering angle and 
$P_l(z)$ the Legendre polynomials
[\ref{GR}].
Below the inelastic threshold, unitarity implies
\equation{
  t_l={1\over2ik}\left(\rme^{2i\delta_l}-1\right),
  \enum
}
with $\delta_l$ the (real) scattering phase 
for angular momentum $l$.

\appendix B

For all $q\geq0$ the angle $\phi(q)$ is determined through
\equation{
  \tan\phi(q)=-{\pi^{3/2}q\over{\cal Z}_{00}(1;q^2)},
  \qquad
  \phi(0)=0,
  \enum
}
and the requirement that it depends continuously on $q$. 
The zeta function in this 
equation is defined by
\equation{
  {\cal Z}_{00}(s;q^2)={1\over\sqrt{4\pi}}
  \sum_{{\bf n}\in\gz^3}({\bf n}^2-q^2)^{-s}
  \enum
}
if $\Re s>\frac{3}{2}$ and elsewhere through analytic continuation.

Numerical methods to compute the zeta function are
described in ref.~[\ref{TwoParticleStates}]
and a table of values of $\phi(q)$ is included in ref.~[\ref{Resonances}]. 
The source code of a set of ANSI~C programs for these 
functions can be obtained from the authors.

% List of references

\ninepoint

\beginbibliography

% Strategies to compute kaon decay rates

\bibitem{DawsonEtAl}
C. Dawson, G. Martinelli, G. C. Rossi, C. T. Sachrajda,
S. Sharpe, M. Talevi, M. Testa,
Nucl. Phys. B514 (1998) 313

% Maiani-Testa "No-go theorem"

\bibitem{MaianiTesta}
L. Maiani, M. Testa,
Phys. Lett. B245 (1990) 585

\bibitem{CiuchiniEtAl}
M. Ciuchini, E. Franco, G. Martinelli, L. Silvestrini,
Phys. Lett. B380 (1996) 353

% Computation of two-particle energies using numerical simulations 

\bibitem{MontvayWeisz}
I. Montvay, P. Weisz,
Nucl. Phys. B290 [FS20] (1987) 327

\bibitem{FrickEtAl}
Ch. Frick, K. Jansen, J. Jers\'ak, I. Montvay, G. M\"unster, P. Seuferling,
Nucl. Phys. B331 (1990) 515

\bibitem{LuscherWolff}
M. L\"uscher, U. Wolff,
Nucl. Phys. B339 (1990) 222

\bibitem{GuagnelliEtAl}
M. Guagnelli, E. Marinari, G. Parisi,
Phys. Lett. B240 (1990) 188

\bibitem{GattringerEtAl}
C. R. Gattringer, C. B. Lang,
Nucl. Phys. B391 (1993) 463

\bibitem{GuptaEtAl}
R. Gupta, A. Patel, S. Sharpe,
Phys. Rev. D48 (1993) 388

\bibitem{FiebigEtAl}
H. R. Fiebig, A. Dominguez, R. M. Woloshyn,
Nucl. Phys. B418 (1994) 649

\bibitem{GoeckelerEtAl}
M. G\"ockeler, H. A. Kastrup, J. Westphalen, F. Zimmermann,
Nucl. Phys. B425 (1994) 413

\bibitem{FukugitaEtAl}
M. Fukugita, Y. Kuramashi, M. Okawa, H. Mino, A. Ukawa,
Phys. Rev. D52 (1995) 3003

\bibitem{AokiEtAl}
S. Aoki et al. (JLQCD Collab.),
Phys. Rev. D58 (1998) 054503

\bibitem{GutsfeldEtAl}
C. Gutsfeld, H. A. Kastrup, K. Stergios,
Nucl. Phys. B560 (1999) 431

% Scattering phases from energy levels in finite volume

\bibitem{StableStates}
M. L\"uscher,
Commun. Math. Phys. 104 (1986) 177

\bibitem{ScatteringStates}
M. L\"uscher,
Commun. Math. Phys. 105 (1986) 153

\bibitem{TwoParticleStates}
M. L\"uscher, 
Nucl. Phys. B354 (1991) 531

\bibitem{Resonances}
M. L\"uscher, 
Nucl. Phys. B364 (1991) 237

\bibitem{RummukainenGottlieb}
K. Rummukainen, S. Gottlieb,
Nucl. Phys. B450 (1995) 397

% Pion scattering phases from one-loop chiral perturbation theory 

\bibitem{GasserLeutwyler}
J. Gasser, H. Leutwyler,
Phys. Lett. B125 (1983) 325;
Ann. Phys. (NY) 158 (1984) 142;
Nucl. Phys. B250 (1985) 465

\bibitem{GasserMeissner}
J. Gasser, U.-G. Meissner,
Phys. Lett. B258 (1991) 219

\bibitem{KnechtEtAl}
M. Knecht, B. Moussallam, J. Stern, N. H. Fuchs,
Nucl. Phys. B457 (1995) 513

% Recent discussion of the influence of final state interactions

\bibitem{PallantePich}
E. Pallante, A. Pich,
Strong enhancement of $\epsilon'/\epsilon$ through final state
interactions,
hep-ph/9911233

\bibitem{Paschos}
E. A. Paschos,
Rescattering effects for $\epsilon'/\epsilon$,
hep-ph/9912230

\bibitem{BurasEtAl}
A. J. Buras, M. Ciuchini, E. Franco, G. Isidori, G. Martinelli, 
L. Silvestrini,
Final state interactions and $\epsilon'/\epsilon$\kern1pt: a critical look,
hep-ph/0002116

% Two-pion states in the quenched approximation

\bibitem{BernardGolterman}
C. W. Bernard, M. F. L. Golterman,
Phys. Rev. D53 (1996) 476

\bibitem{GoltermanLeung}
M. F. L. Golterman, K. C. Leung,
Phys. Rev. D56 (1997) 2950;
{\it ibid.} D57 (1998) 5703;
{\it ibid.} D58 (1998) 097503

% Table of integrals, ...

\bibitem{GR}
I. S. Gradshteyn, I. M. Ryzhik,
Table of Integrals, Series and Products
(Academic Press, New York, 1965) 

\endbibliography

\bye